\documentclass[preprint,12pt]{aastex}

\shortauthors{Winn et al.~2008}
\shorttitle{Albedo of TrES-3}

\begin{document}

%
\def\ltsima{$\; \buildrel < \over \sim \;$}
\def\lsim{\lower.5ex\hbox{\ltsima}}
\def\gtsima{$\; \buildrel > \over \sim \;$}
\def\gsim{\lower.5ex\hbox{\gtsima}}
\def\lam{\lambda=-1\fdg4 \pm 1\fdg1}
%

\bibliographystyle{apj}

\title{
The Transit Light Curve Project.\\
VIII.~Six Occultations of the Exoplanet TrES-3
}

\author{
Joshua N.\ Winn\altaffilmark{1},
Matthew J.\ Holman\altaffilmark{2},
Avi Shporer\altaffilmark{3},
Jos\'e Fern\'andez\altaffilmark{2},\\
Tsevi Mazeh\altaffilmark{3},
David W.\ Latham\altaffilmark{2},
David Charbonneau\altaffilmark{2},
Mark E.~Everett\altaffilmark{4}
}

\altaffiltext{1}{Department of Physics, and Kavli Institute for
  Astrophysics and Space Research, Massachusetts Institute of
  Technology, Cambridge, MA 02139, USA}

\altaffiltext{2}{Harvard-Smithsonian Center for Astrophysics, 60
  Garden Street, Cambridge, MA 02138, USA}

\altaffiltext{3}{Wise Observatory, Tel Aviv University, Tel Aviv
  69978, Israel}

\altaffiltext{4}{Planetary Science Institute, 1700 E.~Fort Lowell Rd.,
  Suite 106, Tucson, AZ 85719}

\begin{abstract}

  We present photometry of the exoplanet host star TrES-3 spanning six
  occultations (secondary eclipses) of its giant planet. No flux
  decrements were detected, leading to 99\%-confidence upper limits on
  the planet-to-star flux ratio of $2.4 \times 10^{-4}$, $5.0\times
  10^{-4}$, and $8.6\times 10^{-4}$ in the $i$, $z$, and $R$ bands
  respectively. The corresponding upper limits on the planet's
  geometric albedo are 0.30, 0.62, and 1.07. The upper limit in the
  $i$ band rules out the presence of highly reflective clouds, and is
  only a factor of 2--3 above the predicted level of thermal radiation
  from the planet.

\end{abstract}


\section{Introduction}

The detection of reflected light from close-in giant planets is a
difficult but worthwhile goal. Knowledge of the albedo is important
because the stellar insolation is a critical factor in the atmospheric
structure and the overall thermal balance of these planets (Guillot et
al.~1996, Saumon et al.~1996, Seager \& Sasselov~1998). Measurement of
the planet's reflection spectrum, or at least the broad
wavelength-dependence of the albedo, would provide clues about the
dominant scattering mechanisms and constituents in the planetary
atmosphere (see, e.g., Seager \& Sasselov 1998, Marley et al.~1999,
Sudarsky et al.~2000, Barman et al.~2001). The difficulty is that the
reflected light is a very small fraction of the direct starlight,
producing a planet-to-star flux ratio at opposition of
$\epsilon_\lambda = p_\lambda (R_p/a)^2$, where $R_p$ is the planetary
radius, $a$ is the orbital separation, and $p_\lambda$ is the
wavelength-dependent geometric albedo.\footnote{The geometric albedo
  is defined as the flux reflected by the planet when viewed at
  opposition (full phase), divided by the flux that would be reflected
  by a flat and perfectly diffusing surface with the same
  cross-sectional area as the planet.}  Even for a planet as large as
Jupiter, and an orbital separation as small as 0.05~AU,
$\epsilon_\lambda \approx 10^{-4}$~$p_\lambda$. The reflected signal
is also expected to vary over the orbital period, in a manner
depending on the orbital inclination and the phase function of the
planetary atmosphere.

Early attempts to detect reflected light from close-in giant planets,
by Charbonneau et al.~(1999), Collier Cameron et al.~(1999, 2002), and
Leigh et al.~(2003a,b), relied on optical spectroscopy. Those
investigators sought the reflected copies of the stellar spectral
lines by combining suitably Doppler-shifted spectra taken over a range
of orbital phases. They did not detect the reflected signal and placed
upper limits\footnote{Although Collier Cameron et al.~(1999) reported
  a detection with greater than 95\% confidence, additional data
  obtained by the same group did not confirm the detection; see
  Collier Cameron et al.~(2002).} on the geometric albedo in the
visual band of $\sim$0.1--0.5, subject to some ambiguity because of
the unknown radius and orbital inclination of the planets they
observed. Liu et al.~(2008) recently revisited the spectroscopic
technique and applied it to HD~209458b, for which the planetary radius
and orbital inclination are known, but the available data provided
only a weak constraint on the geometric albedo ($0.8\pm 1.6$ from
554$-$681~nm).

Another detection method involves polarimetry. The Stokes parameters
are expected to vary over the planetary orbital period because
reflected light is preferentially polarized while direct starlight is
unpolarized (see, e.g., Hough et al.~2006). Berdyugina et al.~(2007)
reported a detection of a time-variable polarized signal from
HD~189733. Taking the scattering radius to be the same as the optical
radius measured through transit photometry, the implied geometric
albedo is larger than $2/3$, which is the geometric albedo of a
perfectly diffusing sphere. While this is physically possible, and
indeed some Solar system objects have geometric albedos exceeding
unity due to strong backscattering, Berdyugina et al.~(2007) preferred
an interpretation in which the albedo is smaller and the scattering
radius is larger than the optical radius measured during transits.

For planets whose orbits are viewed close enough to edge-on that they
undergo periodic occultations by the star, a powerful and conceptually
simple method is available: when the planet is hidden by the star, the
total light should diminish by the fraction
$\epsilon_\lambda$. Edge-on systems are also advantageous because the
planetary radius and orbital inclination can be determined precisely
from transit observations.\footnote{It is possible, however, for a
  planet on an eccentric orbit to exhibit occultations without
  transits, or transits without occultations (see, e.g., Irwin et
  al.~2008).} If the measurement noise were limited only by photon
statistics, the photometric technique could be employed with a smaller
telescope than would be needed for either the spectroscopic or
polarimetric techniques. However, ground-based photometry is generally
afflicted by systematic errors that prevent one from achieving the
required precision. To date, only spaceborne photometry has provided
meaningful upper limits on exoplanetary albedos. In particular, Rowe
et al.~(2007) have used the {\it Microvariability and Oscillations of
  Stars}\, satellite to set a 3$\sigma$ upper limit of 0.17 on the
geometric albedo of HD~209458.

Discoveries of transiting planets have abounded in the past few years,
and there are now several systems known with larger values of
$(R_p/a)^2$. One of the most favorable systems is TrES-3, discovered
by O'Donovan et al.~(2007), for which $(R_p/a)^2 = 7.5\times
10^{-4}$. In this system, a planet with mass 2~$M_{\rm Jup}$ and
radius 1.3~$R_{\rm Jup}$ circles a G dwarf star with an orbital period
of 31~hr. This paper relates our attempts to detect reflected light
from the TrES-3 planet using ground-based photometry with meter-class
telescopes. In \S~2, we describe our observations and data reduction
procedures.  In \S~3, we present our results. In \S~4, we place these
results in the context of other observations and of theoretical
expectations for the albedo and the thermal emission spectrum.

\section{Observations and Data Reduction}

We observed TrES-3 on 6 different nights when planetary occultations
were expected, according to the ephemeris of O'Donovan et
al.~(2007). Assuming that the planetary orbit is circular, as expected
for a planet with such a short orbital period\footnote{The timescale
  for tidal circularization is $\sim$$2\times 10^7$~yr, using Eqn.~(9)
  of Rasio et al.~(1996) with the system parameters of O'Donovan et
  al.~(2007) and a tidal quality factor $Q=10^5$. This is shorter than
  the estimated age of the star, $5\times 10^8$~yr (Torres et
  al.~2008). The current radial velocity data are consistent with a
  circular orbit and give an upper limit on the eccentricity of 0.12
  with 99\% confidence.}, the uncertainty in the predicted
mid-occultation time was smaller than 3 minutes on each night.
Observations on other nights were attempted, but only the six nights
described below offered clear enough skies for high-precision
photometry.

On 2007~April~27, 2007~July~4, 2007~September~14, and 2008~March~12,
we used the 1.2~m telescope at the Fred L.\ Whipple Observatory (FLWO)
on Mount Hopkins, Arizona. (Here and elsewhere, the quoted date refers
to the UT date at the start of the night.) The detector was KeplerCam,
a 4096$^2$ CCD with a square field of view $23\farcm 1$ on a side
(Szentgyorgi et al.~2005). We binned the images $2\times 2$, giving a
scale of $0\farcs68$ per binned pixel. We obtained repeated 60~s
exposures over the course of 4-5~hr bracketing the predicted
mid-occultation time. The dead time between exposures was
11~s. Autoguiding maintained the pointing to within 2-3 pixels
throughout the observations. On the first night, 2007~April~27, we
observed through a Sloan $i$ filter ($\approx$0.7--0.85~$\mu$m), as
the star rose from an airmass of 1.4 to the meridian. The full-width
at half-maximum (FWHM) of stellar images was about 4.5~pixels
($3\arcsec$). On the next two nights, 2007~July~4 and
2007~September~14, we observed through a Sloan $z$ filter
($\approx$0.85--1.0~$\mu$m, with the red cutoff arising from the
quantum efficiency of the detector). On both of those nights, the
target star began near the meridian and set to an airmass near 2.0
throughout the observations, and the FWHM was approximately 3~pixels
($2\arcsec$). On the last night, 2008~March~12, we chose the Sloan $i$
filter again and observed through an airmass ranging from 1.1 to 2.0.
Conditions were nearly photometric, and the seeing was exceptionally
good and stable, with a FWHM of 2.1~pixels ($1\farcs4$).

On UT~2007~August~26 and 2007~October~3, we used the 1~m telescope at
Wise Observatory in Israel. We used a Princeton Instruments
$1340\times 1300$ back-illuminated CCD, giving a field of view of
$13\farcm 0 \times 12\farcm 6$ and a pixel scale of $0\farcs 58$. The
readout and refresh time was 25~s, and we used exposure times ranging
from 60~s to 160~s, depending on the airmass and seeing. The guider
generally maintained the pointing to within a few pixels, but there
was a glitch on each night that led to a 10-20 pixel offset. We
observed through a Bessell $R$ filter ($\approx$0.55--0.7~$\mu$m). On
August~26, we observed for 6~hr as the target star set from the
meridian to an airmass of 2.8, and on October~3, we observed for 4~hr
as the target set from airmass 1.1 to 2.5. In both cases, the data
obtained through an airmass greater than 2.0 proved to be much noisier
than the rest of the data, and ultimately were not used in our
analysis. On both nights the seeing steadily worsened with increasing
airmass, from a FWHM of 2.5~pixels to 4.5~pixels ($1\farcs8$ to
$3\farcs2$).

We used standard IRAF procedures for overscan correction, trimming,
bias subtraction, and flat-field division. We performed aperture
photometry of TrES-3 and several nearby stars. The sum of the fluxes
of the comparison stars was taken to be the comparison signal. The
flux timeseries for TrES-3 was divided by the comparison signal, and
then by a constant chosen to give a unit mean flux outside of the
predicted occultation. We experimented with different choices for the
aperture size, and different combinations of the comparison stars,
aiming to minimize the standard deviation of the out-of-occultation
portion of the TrES-3 light curve. Not surprisingly, the best results
were obtained when the aperture diameter was about twice as large as
the FWHM of the stellar images, and when as many comparison stars as
possible were used with mean fluxes ranging from 50--150\% of the mean
flux of TrES-3. For the FLWO data, 10-13 stars were used, and for the
Wise data, 6-7 stars were used. The Wise light curve had an abrupt
jump at precisely the time when the guiding failed and the pointing
changed; to correct for this, we set the mean out-of-occultation flux
to be unity for each of the two segments of data. Finally, for all
data sets, we performed an airmass correction. Using only the
out-of-occultation data, we fitted the light curve to an exponential
function of airmass, divided by the best-fitting function, and
renormalized the light curve to have unit mean flux outside of the
occultation.

The final relative photometry is plotted in Figure~1, and are given in
numerical form in Table~1. One of the transit light curves of
O'Donovan et al.~(2007) is also plotted in Figure~1, because the
transit data were used in our model for the occultation data, as
described in the next section. The duration of the occultation is
expected to be nearly the same as the duration of the transit, since
the orbit is presumed to be circular, as discussed further in \S~4.

\begin{figure}[p]
\epsscale{0.9}
\plotone{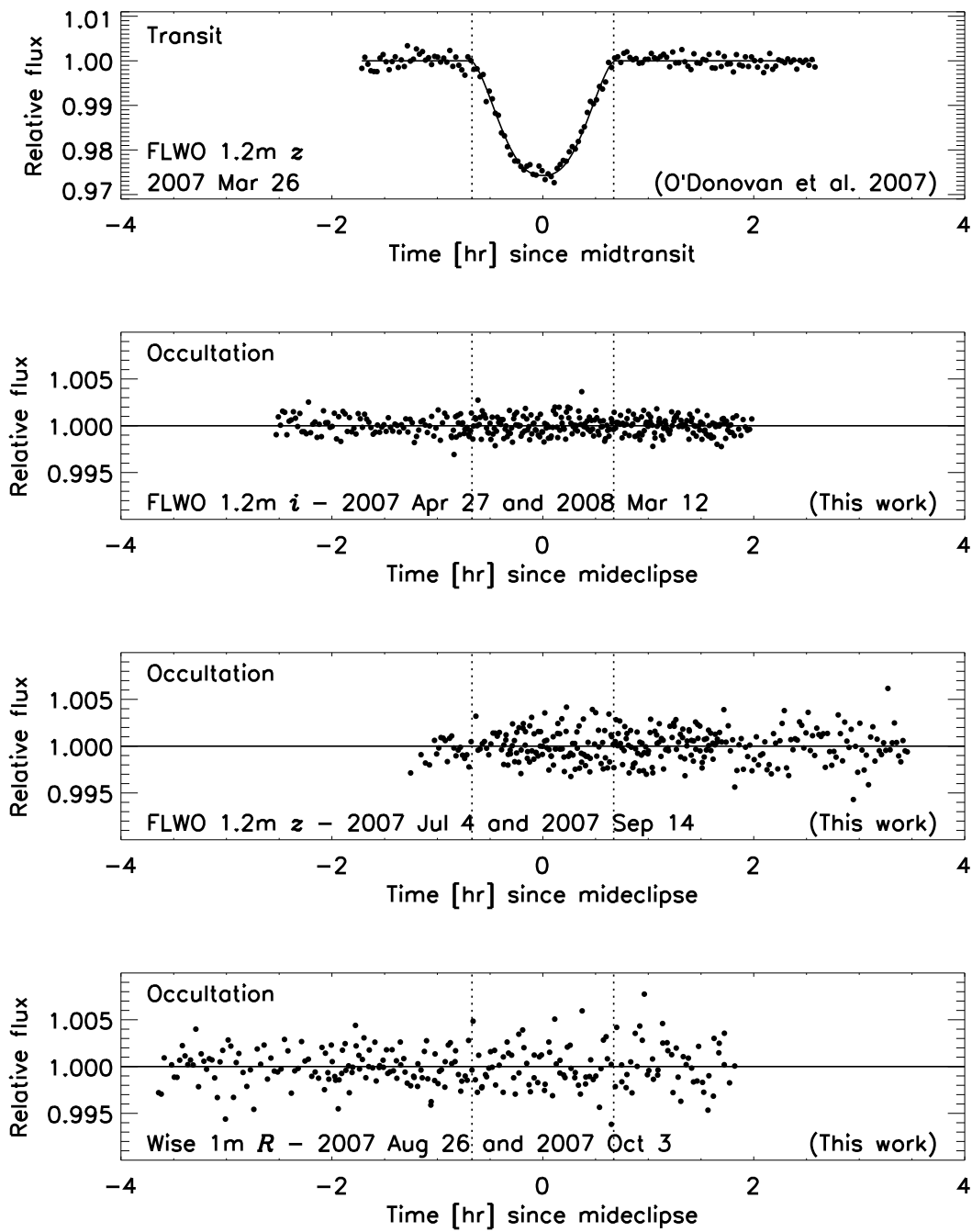}
\caption{Photometry of TrES-3. The top panel shows the $z$-band
  transit photometry of O'Donovan et al.~(2007). The other 3 panels
  show our photometry spanning occultations (secondary eclipses).
  The dotted lines show the expected start and end of the occultation.
  \label{fig:1}}
\end{figure}

\begin{deluxetable}{lccc}
\tabletypesize{\normalsize}
\tablecaption{Photometry of TrES-3\label{tbl:photometry}}
\tablewidth{0pt}

\tablehead{
\colhead{Telescope} &
\colhead{Filter} &
\colhead{Heliocentric Julian Date} & 
\colhead{Relative flux}
}

\startdata
      FLWO &     i &  2454217.806989 &    0.99905 \\
      FLWO &     i &  2454217.807834 &    1.00095 \\
      FLWO &     i &  2454217.808667 &    1.00006 \\
      FLWO &     i &  2454217.809524 &    1.00156 \\
      FLWO &     i &  2454217.810369 &    1.00144 \\
      FLWO &     i &  2454217.811214 &    0.99904
\enddata 

\tablecomments{The time stamps represent the Heliocentric Julian Date
  at the time of mid-exposure. We intend for this Table to appear in
  entirety in the electronic version of the journal. An excerpt is
  shown here to illustrate its format. The data are also available
  from the authors upon request.}

\end{deluxetable}

Since no flux decrement is obvious in any of the light curves, the
results are upper limits on the planet-to-star flux ratio in each
bandpass. This makes it especially important to understand the
characteristics of the noise. The out-of-occultation FLWO $i$ light
curves from 2007~April~26 and 2008~March~12 have standard deviations
of $9.1\times 10^{-4}$ and $9.0\times 10^{-4}$, respectively. The
expected level of photon-counting (Poisson) noise is $8\times
10^{-4}$, and the expected level of scintillation noise is $3\times
10^{-5}$, according to Eqn.~1 of Young~(1967). The quadrature sum of
these terms is $8.5\times 10^{-4}$. The FLWO $z$ light curves from
2007~July~14 and 2007~September~14 have standard deviations of
$1.8\times 10^{-3}$ and $1.4\times 10^{-3}$, as compared to the
expected level of about $1.2\times 10^{-3}$ from Poisson and
scintillation noise. Thus, most of the observed noise can be accounted
for. The noise in the $i$-band light curves, in particular, is only
6-7\% larger than the noise from these two expected sources.

The out-of-occultation Wise data from 2007~August~26 and
2007~October~3 have standard deviations of $2.2\times 10^{-3}$ and
$1.9\times 10^{-3}$. In both cases the expected level of
photon-counting noise is about $10^{-3}$ and the expected level of
scintillation noise is about $3\times 10^{-5}$. Thus, the noise in the
Wise data is about twice as large as one would expect from these two
sources. Because of the jump that was observed in each light curve at
the time of a large pixel shift, we suspect that at least some (and
perhaps most) of the excess noise is due to an imperfect flat
field. On other nights when the telescope lost tracking more
frequently the noise level was indeed even higher, although those
other nights were also characterized by poorer sky conditions, making
it impossible to isolate the source of excess noise.

Next we tested for any correlations with time or other external
variables, and for significant non-Gaussianity. We found no
significant correlation (a correlation coefficient $<$0.2) between the
final relative flux data and the airmass, the pixel position, or the
measured shape parameters (FWHM, ellipticity, position angle) of
stellar images. Of course, we had already decorrelated against the
airmass, and for the Wise data we had already removed the effects of
the two pointing glitches, so the lack of correlations with those
variables was expected. For each light curve, we also examined the
histogram of flux values, the autocorrelation function, and the factor
by which the noise level is reduced when the data are binned. Because
the histograms are approximately Gaussian, the autocorrelations are
generally as small as would be expected for uncorrelated noise, and
the noise falls approximately as $1/\sqrt{N}$, in what follows we
treat the noise as Gaussian and uncorrelated.

\section{Determination of Upper Limits on the Albedo}

One might imagine measuring the planet-to-star flux ratio,
$\epsilon_\lambda$, by finding the ratio of the mean flux during the
occultation to the mean out-of-occultation flux. However, because the
TrES-3 planet is on a nearly-grazing trajectory, the durations of the
ingress and egress cannot be ignored. Furthermore, the uncertainty in
the orbital inclination $i$ introduces some uncertainty in the
expected shape of the light curve, and the conversion from
$\epsilon_\lambda$ into the geometric albedo $p_\lambda$ requires
knowledge of the ratio $R_p/a$ and its uncertainty.

To take all of these factors into account, we simultaneously fitted a
parameterized model to our $R$, $i$, and $z$-band occultation
photometry and the high-precision $B$ and $z$-band transit photometry
of O'Donovan et al.~(2007). The model and the fitting method are
similar to those we have employed in previous papers in this series
(see, e.g., Winn et al.~2007, Holman et al.~2007). The model posits a
circular orbit of a planet with radius $R_p$ and a star with radius
$R_\star$, with orbital inclination $i$, orbital separation $a$, and
period $P$. The flux ratio between the full disk of the star and
fully-illuminated disk of the planet is $\epsilon_k$, where $k$ refers
to the bandpass. When the planet is projected in front of the star, or
vice versa, we use the analytic formulae of Mandel \& Agol (2002) to
compute the appropriate flux decrement. We assumed that the flux
received from the planet does not vary appreciably over the 5~hr span
of our occultation data, i.e., that the planet's phase function is
constant within $30\arcdeg$ of opposition. For the star, we assumed
the limb-darkening law to be quadratic, with
coefficients\footnote{Specifically, we used $a=0.2508$,~$b=0.3019$ for
  the $z$-band data; and $a=0.7250$,~$b=0.0967$ for the $B$-band
  data. These are based on interpolations for a star with $T_{\rm
    eff}=5720$~K, $\log g=4.6$, and solar
  metallicity. Southworth~(2008) has found that the uncertainties in
  the transit parameters (the scaled radii of the star and planet, and
  the orbital inclination) are underestimated when the limb-darkening
  coefficients are fixed at theoretical values. However, our results
  for the geometric albedo are not susceptible to this problem, as we
  confirmed by repeating the fit with a linear limb-darkening law and
  allowing the coefficients to be free parameters.} taken from
Claret~(2000, 2004). The model parameters were $\epsilon_i$,
$\epsilon_z$, $\epsilon_R$, $R_p/a$, $R_\star/a$, $i$, $P$, and $T_c$
(a particular time of midtransit), along with the airmass correction
parameters for each light curve.

For the transit light curves, we adopted the same flux uncertainties
as O'Donovan et al.~(2007). For the occultation light curves, based on
the investigation of the photometric errors described in the previous
section, we took the standard deviation of the out-of-occultation flux
to be the uncertainty in each flux measurement. To determine the
best-fitting values of the parameters and their uncertainties, we used
a Markov Chain Monte Carlo (MCMC) algorithm. This algorithm creates a
random sequence of points in parameter space whose density is an
approximation of the {\it a posteriori}\, probability distribution of
the parameter values. The sequence is generated from an initial point
by iterating a ``jump function,'' which in our case was the addition
of a Gaussian random number to a randomly-chosen parameter. If the new
point has a lower $\chi^2$ than the previous point, the jump is
executed; if not, the jump is executed with probability
$\exp(-\Delta\chi^2/2)$. We used an isotropic prior for the orbital
inclination (uniform in $\cos i$), and a Gaussian prior in the orbital
period to enforce consistency with the value $P=1.30619 \pm
0.00001$~d, which was determined by O'Donovan et al.~(2007) using the
yearlong baseline between the TrES survey data and the high-precision
light curves. We used uniform priors for all other parameters. The
planet-to-star flux ratios were required to be positive. After
creating several independent chains to verify that they all converged
to the same region of parameter space, we created one long chain of
$5\times 10^6$ points for our final results.

In Table~2, we provide the results for the 99\%-confidence upper limit
on the planet-to-star flux ratio and the geometric albedo for each
bandpass. The geometric albedo was computed as $p_\lambda =
\epsilon_\lambda(a/R_p)^2$. The upper limit was defined as the value
for which the cumulative {\it a posteriori}\, probability distribution
took the value 0.99. Figure~2 shows the {\it a posteriori}\,
probability distribution for the geometric albedo based on the data
from each bandpass, and for all of the data in combination. The
results for the transit-based parameters $R_p/R_\star$, $R_\star/a$,
$i$, and $T_c$ were in agreement with the published values of
O'Donovan et al.~(2007).

\begin{figure}[ht]
\epsscale{0.5}
\plotone{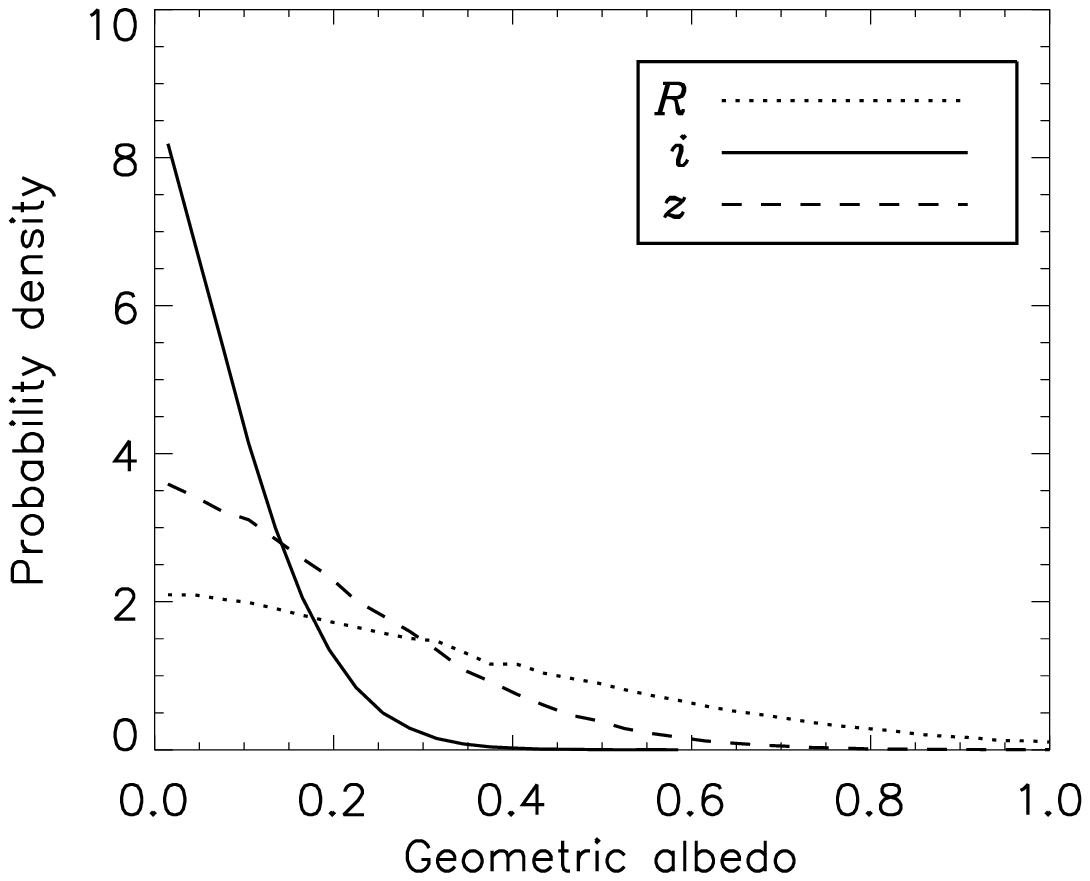}
\caption{Constraints on the geometric albedo of exoplanet TrES-3, averaged
over various bandpasses. Each curve shows the {\it a posteriori}\,
probability distribution determined with the MCMC algorithm.
\label{fig:2}}
\end{figure}

\begin{deluxetable}{ccc}
\tabletypesize{\normalsize}
\tablecaption{Results of Occultation Photometry of TrES-3\label{tbl:params}}
\tablewidth{0pt}

\tablehead{
\colhead{Bandpass} &
\multicolumn{2}{c}{Upper Limit (99\% confidence)} \\
\colhead{} &
\colhead{Planet-to-star flux ratio} &
\colhead{Geometric albedo}
}

\startdata
$i$   & $0.00024$ & $0.30$ \\
$z$   & $0.00050$ & $0.62$ \\
$R$   & $0.00086$ & $1.07$
\enddata

\end{deluxetable}

\section{Discussion and Summary}

This study has resulted in empirical upper limits on the
planet-to-star flux ratio, and the geometric albedo, in the wavelength
range 0.6$-$1.0~$\mu$m. The most stringent upper limit was obtained in
the Sloan $i$ band, where $p_i < 0.30$ with 99\% confidence. To put
this result in perspective, we remind the reader that a
perfectly-diffusing sphere has a geometric albedo of $2/3$. The TrES-3
planet is less reflective than such a sphere, at least for
observations in the $i$ and $z$ bands. The bandpass-averaged geometric
albedos of Jupiter in the $R$, $i$, and $z$ bands are approximately
0.50, 0.45, and 0.25, respectively (Karkoschka 1994). Our results show
that TrES-3 is less reflective than Jupiter in the $i$ band, although
they do not enable a meaningful comparison in the other two
bands. Firm comparisons to other exoplanets are also difficult at this
stage because almost all published reports are upper limits. Using the
spectroscopic method, Charbonneau et al.~(1999), Collier Cameron et
al.~(2002), and Leigh et al.~(2003a,b) have placed upper limits of
0.1--0.4 on various systems, although in those cases the results are
subject to extra uncertainty because the planetary radius and orbital
inclinations are not known. Rowe et al.~(2007) used several months of
spaceborne photometry of HD~209458 to conclude $p_{\rm opt} <0.17$ for
that system with 3$\sigma$ (99.73\%) confidence, over an optical
bandpass ranging from 0.4--0.7~$~\mu$m. At the same confidence level
of 99.73\%, our upper limit in the Sloan $i$ band is $p_i < 0.35$. As
mentioned in \S~1, Berdyugina et al.~(2007) found a surprisingly large
planet-to-star flux ratio for HD~189733, indeed large enough that
those authors suggested the scattering radius is larger than the
transit radius. Our results do not permit such an interpretation for
TrES-3.

Theoretical reflection spectra for strongly irradiated giant planets
such as TrES-3 depend upon many factors and are especially sensitive
to the presence or absence of clouds. Without highly reflective
clouds, they are generally expected to have very low optical albedos.
The underlying cause of the low albedo in cloud-free models is strong
absorption in the pressure-broadened resonance doublet transitions of
sodium and potassium (see, e.g., Seager et al.~2000, Marley et
al.~1999, Sudarsky et al.~2000) or, for the hottest planets, the
plentiful electronic transitions of gaseous TiO and VO (Hubeny et
al.~2003, Sharp \& Burrows 2007, Fortney et al.~2007). Condensates may
form reflective clouds that can drastically raise the albedo. The
particular condensates that are available will depend on the
temperature and pressure profile of the atmosphere.

As a specific example, Sudarsky et al.~(2000, 2003) predicted that gas
giants with effective temperatures between about 900~K and 1500~K
(``Class IV'' in their terminology) have optical albedos $\lsim 0.2$
in the bandpasses we employed, while for temperatures $\gsim 1500$~K
(``Class V''), the albedo is as large as 0.6. The difference is that
the hotter planets have a high layer of opaque silicate and iron
clouds in their upper atmospheres. The case of TrES-3 is borderline,
with an effective temperature of 1643~$f^{1/4} (1-A)^{1/4}$~K
(O'Donovan et al.~2007), where $A$ is the Bond albedo (the ratio of
reflected to incident bolometric power) and $f$ is a phenomenological
factor accounting for atmospheric circulation (with $f=1$ for
isotropic reradiation and $f=2$ for radiation from the dayside
only). For TrES-3 it would be possible to obtain a self-consistent
Class~V solution with highly-reflective silicate and iron clouds,
e.g., $f=1$ and $A\approx 0.4$, giving $T_{\rm eff}=1500$~K. Our upper
limit on $p_i$ rules this out, and hence may be regarded as evidence
against the presence of highly reflective clouds in the planet's upper
atmosphere.

For cloud-free atmospheres, the thermal radiation from the planet is
expected to produce a larger signal than the reflected light (see,
e.g., L\'opez-Morales \& Seager 2007, Fortney et al.~2007). One might
wonder whether our limits on the planet-to-star flux ratio also
constrain the thermal emission spectrum of the planet. The thermally
emitted flux can be estimated from the effective temperature of the
planet. For TrES-3, even for the hottest possible temperature of
2100~K (for $f=2$ and $A=0$) we estimate that the maximum
planet-to-star flux ratio due solely to thermal emission is $2\times
10^{-4}$ at a wavelength of 0.9~$\mu$m. This rough estimate agrees
within a factor of 2 with a more detailed atmospheric model of TrES-3
by Fortney et al.~(2007). Those investigators classify TrES-3 as a pM
planet\footnote{The distinction between pM and pL planets, in the
  nomenclature proposed by Fortney et al.~(2007), is analogous to the
  M/L transition in low-mass stars and brown dwarfs. The pM planets
  are hot enough for TiO and VO to exist in gaseous form, leading to
  strong absorption of the stellar flux at low pressure, a temperature
  inversion in the planet's upper atmosphere, and a brightness
  temperature in excess of the equilibrium temperature in the optical
  and infrared bands.}, for which they predict a very low albedo
(because conditions are too hot for any condensates) and an
``anomalously hot'' brightness temperature (because of a high-altitude
temperature inversion). For our observing bandpass the calculated
planet-to-star flux ratio is $\sim$$10^{-4}$. This is smaller than our
most constraining upper limit of $2.4\times 10^{-4}$ in the $i$-band,
and hence we conclude that the expected level of thermal emission is
beneath our detection limit by a factor of 2--3.

In summary, through high-precision photometry of an especially
favorable target, we have placed meaningful upper limits on the albedo
of a transiting exoplanet. This is the first time this has been done
with ground-based occultation photometry. Our results seem to be
limited by random noise, suggesting that the acquisition of more data
will lead to increased sensitivity.  For example, with 8 more light
curves comparable in quality to the two $i$-band light curves
presented here, the photon-limited $1\sigma$ error in the albedo would
be approximately 0.03. Even if the albedo proves to be very small, as
expected in cloud-free atmospheric models, such a data set would
enable a $\sim$4$\sigma$ detection of the expected level of thermal
emission.  Of course, it is possible that sources of systematic noise
will become limiting factors, such as flat-fielding errors, or
time-variable differential extinction beyond a gradual airmass
dependence. Nevertheless, our results give reason to hope for an
unambiguous detection of reflected light or thermal emission from
exoplanets based on ground-based photometry using meter-class
telescopes.

\acknowledgments We thank Jonathan Fortney, Willie Torres, and Alex
Sozzetti for helpful conversations and correspondence; John
Southworth, for his code for computing limb-darkening coefficients;
and an anonymous referee for a timely and detailed review. This
research was supported by Grant No.~2006234 from the United
States-Israel Binational Science Foundation (BSF), Jerusalem,
Israel. KeplerCam was developed with partial support from the {\it
  Kepler}\, mission under NASA Cooperative Agreement NCC2-1390
(PI:~D.~Latham).


\begin{thebibliography}{}
 
\bibitem[Barman et al.(2001)]{2001ApJ...556..885B} Barman, T.~S.,
  Hauschildt, P.~H., \& Allard, F.\ 2001, \apj, 556, 885

\bibitem[Berdyugina et al.(2008)]{2008ApJ...673L..83B} Berdyugina,
  S.~V., Berdyugin, A.~V., Fluri, D.~M., \& Piirola, V.\ 2008, \apjl,
  673, L83

\bibitem[Charbonneau et al.(1999)]{1999ApJ...522L.145C} Charbonneau,
  D., Noyes, R.~W., Korzennik, S.~G., Nisenson, P., Jha, S., Vogt,
  S.~S., \& Kibrick, R.~I.\ 1999, \apjl, 522, L145

\bibitem[Claret(2000)]{2000A&A...363.1081C} Claret, A.\ 2000, \aap,
  363, 1081

\bibitem[Claret(2004)]{2004A&A...428.1001C} Claret, A.\ 2004, \aap,
  428, 1001
 
\bibitem[Collier Cameron et al.(1999)]{1999Natur.402..751C} Cameron,
  A.~C., Horne, K., Penny, A., \& James, D.\ 1999, \nat, 402, 751

\bibitem[Collier Cameron et al.(2002)]{2002MNRAS.330..187C} Collier
  Cameron, A., Horne, K., Penny, A., \& Leigh, C.\ 2002, \mnras, 330,
  187

\bibitem[Fortney et al.(2007)]{2007arXiv0710.2558F} Fortney, J.~J.,
  Lodders, K., Marley, M.~S., \& Freedman, R.~S.\ 2007, ArXiv
  e-prints, 710, arXiv:0710.2558
 
\bibitem[Guillot et al.(1996)]{1996ApJ...459L..35G} Guillot, T.,
  Burrows, A., Hubbard, W.~B., Lunine, J.~I., \& Saumon, D.\ 1996,
  \apjl, 459, L35
 
\bibitem[Holman et al.(2007)]{2007ApJ...664.1185H} Holman, M.~J., et
  al.\ 2007, \apj, 664, 1185
 
\bibitem[Hough et al.(2006)]{2006SPIE.6269E..25H} Hough, J.~H., Lucas,
  P.~W., Bailey, J.~A., Tamura, M., \& Hirst, E.\ 2006, \procspie,
  6269, 62690S-2

\bibitem[Hubeny et al.(2003)]{2003ApJ...594.1011H} Hubeny, I.,
  Burrows, A., \& Sudarsky, D.\ 2003, \apj, 594, 1011
 
\bibitem[Irwin et al.(2008)]{2008arXiv0801.1496I} Irwin, J., et al.\
  2008, ArXiv e-prints, 801, arXiv:0801.1496

\bibitem[Karkoschka(1994)]{1994Icar..111..174K} Karkoschka, E.\ 1994,
  Icarus, 111, 174

\bibitem[Leigh et al.(2003a)]{2003MNRAS.346L..16L} Leigh, C., Collier
  Cameron, A., Udry, S., Donati, J.-F., Horne, K., James, D., \&
  Penny, A.\ 2003a, \mnras, 346, L16
 
\bibitem[Leigh et al.(2003b)]{2003MNRAS.344.1271L} Leigh, C., Cameron,
  A.~C., Horne, K., Penny, A., \& James, D.\ 2003b, \mnras, 344, 1271

\bibitem[Liu et al.(2008)]{2007arXiv0711.2304L} Liu, X., et al.\ 2008,
  ArXiv e-prints, 711, arXiv:0711.2304
 
\bibitem[L{\'o}pez-Morales \& Seager(2007)]{2007ApJ...667L.191L}
  L{\'o}pez-Morales, M., \& Seager, S.\ 2007, \apjl, 667, L191

\bibitem[Mandel \& Agol(2002)]{2002ApJ...580L.171M} Mandel, K., \&
  Agol, E.\ 2002, \apjl, 580, L171
 
\bibitem[Marley et al.(1999)]{1999ApJ...513..879M} Marley, M.~S.,
  Gelino, C., Stephens, D., Lunine, J.~I., \& Freedman, R.\ 1999,
  \apj, 513, 879

\bibitem[O'Donovan et al.(2007)]{2007ApJ...663L..37O} O'Donovan,
  F.~T., et al.\ 2007, \apjl, 663, L37

\bibitem[Rasio et al.(1996)]{1996ApJ...470.1187R} Rasio, F.~A., Tout,
  C.~A., Lubow, S.~H., \& Livio, M.\ 1996, \apj, 470, 1187

\bibitem[Rowe et al.(2007)]{2007arXiv0711.4111R} Rowe, J.~F., et al.\
  2007, ArXiv e-prints, 711, arXiv:0711.4111
 
\bibitem[Saumon et al.(1996)]{1996ApJ...460..993S} Saumon, D.,
  Hubbard, W.~B., Burrows, A., Guillot, T., Lunine, J.~I., \&
  Chabrier, G.\ 1996, \apj, 460, 993

\bibitem[Seager \& Sasselov(1998)]{1998ApJ...502L.157S} Seager, S., \&
  Sasselov, D.~D.\ 1998, \apjl, 502, L157
 
\bibitem[Seager et al.(2000)]{2000ApJ...540..504S} Seager, S.,
  Whitney, B.~A., \& Sasselov, D.~D.\ 2000, \apj, 540, 504
 
\bibitem[Sharp \& Burrows(2007)]{2007ApJS..168..140S} Sharp, C.~M., \&
  Burrows, A.\ 2007, \apjs, 168, 140
 
\bibitem[Southworth(2008)]{2008arXiv0802.3764S} Southworth, J.\ 2008,
  ArXiv e-prints, 802, arXiv:0802.3764

\bibitem[Sudarsky et al.(2000)]{2000ApJ...538..885S} Sudarsky, D.,
  Burrows, A., \& Pinto, P.\ 2000, \apj, 538, 885

\bibitem[Sudarsky et al.(2003)]{2003ApJ...588.1121S} Sudarsky, D.,
  Burrows, A., \& Hubeny, I.\ 2003, \apj, 588, 1121
 
\bibitem[Szentgyorgyi et al.(2005)]{2005AAS...20711010S} Szentgyorgyi,
  A.~H., et al.\ 2005, Bulletin of the American Astronomical Society,
  37, 1339
 
\bibitem[Torres et al.(2008)]{2008arXiv0801.1841T} Torres, G., Winn,
  J.~N., \& Holman, M.~J.\ 2008, ArXiv e-prints, 801, arXiv:0801.1841

\bibitem[Winn et al.(2007)]{2007AJ....134.1707W} Winn, J.~N., et al.\
  2007, \aj, 134, 1707
 
\bibitem[Young(1967)]{1967AJ.....72..747Y} Young, A.~T.\ 1967, \aj,
  72, 747
 
\end{thebibliography}
\end{document}